\documentclass[%
reprint,
superscriptaddress,
 aps,
]{revtex4-2}

\usepackage{graphicx}% Include figure files
\usepackage{dcolumn}% Align table columns on decimal point
\usepackage{bm}% bold math
\usepackage{float}
\usepackage{caption}
\usepackage{subcaption}
\usepackage{comment}
\usepackage{amsmath}
\usepackage{hyperref}% add hypertext capabilities
\hypersetup{colorlinks=true, linkcolor=blue, urlcolor=black, filecolor=magenta, citecolor=blue}
\usepackage{csquotes}
\usepackage{gensymb}

\usepackage{lipsum}% http://ctan.org/pkg/lipsum
\usepackage{graphicx}% http://ctan.org/pkg/graphicx

\begin{document}

%\preprint{AIP/123-QED}

\title{Tunable compact on-chip superconducting switch}

\author{Julia Zotova}
\email{yuliya.zotova@phystech.edu}
\affiliation{Skolkovo Institute of Science and Technology, 121205 Moscow, Russia}
\affiliation{Moscow Institute of Physics and Technology, Institutskiy Pereulok 9, Dolgoprudny 141701, Russia}
\affiliation{RIKEN Center for Quantum Computing (RQC), Wako, Saitama 351-0198, Japan}

\author{Alexander Semenov}%
\affiliation{Moscow Institute of Physics and Technology, Institutskiy Pereulok 9, Dolgoprudny 141701, Russia}
\affiliation{Moscow State Pedagogical University, Malaya Pirogovskaya Street 1/1, Moscow 119435, Russia}

\author{Rui Wang}%
\affiliation{Research Institute for Science and Technology, Tokyo University of Science, 1-3 Kagurazaka, Shinjuku-ku, Tokyo 162-8601, Japan}
%\affiliation{Department of Physics, Tokyo University of Science, 1–3 Kagurazaka, Shinjuku, Tokyo 162–0825, Japan}
\affiliation{RIKEN Center for Quantum Computing (RQC), Wako, Saitama 351-0198, Japan}

\author{Yu Zhou}%
\affiliation{RIKEN Center for Quantum Computing (RQC), Wako, Saitama 351-0198, Japan}

\author{Oleg Astafiev}%
\affiliation{Skolkovo Institute of Science and Technology, 121205 Moscow, Russia}
\affiliation{Moscow Institute of Physics and Technology, Institutskiy Pereulok 9, Dolgoprudny 141701, Russia}

\author{Jaw-Shen Tsai}%
\email{tsai@riken.jp}
\affiliation{Research Institute for Science and Technology, Tokyo University of Science, 1-3 Kagurazaka, Shinjuku-ku, Tokyo 162-8601, Japan}
%\affiliation{Department of Physics, Tokyo University of Science, 1–3 Kagurazaka, Shinjuku, Tokyo 162–0825, Japan}
\affiliation{RIKEN Center for Quantum Computing (RQC), Wako, Saitama 351-0198, Japan}

\date{\today}% It is always \today, today,

\begin{abstract}
We develop a compact four-port superconducting switch with a tunable operating frequency in the range of 4.8~GHz -- 7.3~GHz.  Isolation between channel exceeds 20~dB over a bandwidth of several hundred megahertz, exceeding  40~dB at some frequencies.  The footprint of the device is $80\times420~\mu$m. The tunability requires only a global flux bias without either permanent magnets or micro-electromechanical structures. As the switch is superconducting, the heat dissipation during operation is negligible. The device can operate at up to -80~dBm, which is equal to $2.5\times 10^6$ photons at 6 GHz  per microsecond. The device show a possibility to be operated as a beamsplitter with tunable splitting ratio.

\end{abstract}

%\keywords{Transmission lines, tunability, superconducting materials}

\maketitle

\section{Introduction}

Recently, superconducting quantum processors demonstrate rapid progress becoming more and more complex~\cite{kjaergaard2020superconducting, devoret2013superconducting}. To solve computational problems, it is necessary to develop superconducting quantum processors with large number of qubits~\cite{arute2019quantum, wu2021strong, ren2022experimental, gong2021quantum}, which results in large sizes of the processors.
There are several approaches to increase of the on-chip density of elements. Possible approaches with a fixed on-chip configuration include a flip chip~\cite{foxen2017qubit}, through-silicon vias multi-layer chip technology~\cite{yost2020solid}, as well as a quasi-2D \textquote{origami-like} structure~\cite{mukai2020pseudo}. A different approach is a flexible circuit routing system, using \textquote{switch on-off} logic. On-demand routing allows one to calibrate a measurement setup~\cite{ranzani2013two} with fewer input-output channels, allows one to study on-chip transport as an alternative to split gates~\cite{al2013cryogenic} and construct  more complex integrated quantum systems. 

Commercially available mechanical pulse-latched switches based on semiconductors dissipate too much heat~\cite{ranzani2013two} to be compatible with millikelvin temperatures. For the same reason, temperature-activated switches~\cite{el2014low} cannot be used in low-temperature experiments. Current-controlled positive-intrinsic-negative (P-I-N) diode switches, and field-effect transistor (FET) diode switches with normally high insertion loss are also unsuitable for low-temperature experiments at powers of the single-photon level. In contrast, RF micro-electro-mechanical (MEMs) switches~\cite{gong2009study, attar2014low} have low insertion loss, high isolation and near-zero power consumption, as well as small sizes. However, due to the movement of the mechanical parts, this type of switches has a relatively slow switching time (2 -- 40~$\mu$s), requires sophisticated fabrication techniques and is very sensitive to humidity and organic contamination~\cite{rebeiz2001rf}. 

Fast switches, based on nanowires in superconducting and normal states have been demonstrated~\cite{wagner2019demonstration}. It is also compact and broadband, but has high internal losses or parasitic reflections $>$ 5~dB, which may not be sufficient for high-fidelity experiments. 
Also, the cryotron switches~\cite{lowell2016thin}, actuated by external current, show fast switching (less than 200~ns) with high isolation ($\sim 10^{-3}$). However, such devices dissipate Joule heating (40~fW per switching) due to small parasitic resistances in the control line, making them difficult to use it for multi-qubit systems.
The lower loss and higher isolation performance have been achieved with circuits using Josephson junctions and SQUIDs, as tunable inductance elements~\cite{naaman2016chip}.  Some versions of switches based on Josephson junctions can also invert signals~\cite{chapman2016general}. The scheme for multiplexed readout using this type of a switch has been proposed and tested~\cite{chapman2016general}. 

All the discussed switches are single-pole-single-throw. As a result, the use of these types of switches for on-chip routing may be limited, since only one channel is involved. Single-pole-double-throw switches offer greater flexibility in controlling the propagation of the EM-field. Proposals for such devices based on lumped-elements have been demonstrated~\cite{naaman2016chip}. Also, a version using coplanar waveguides and a flux qubit as the coupling element~\cite{baust2015tunable} has been realized experimentally, however, with relatively poor isolation efficiency (62\%) to use it for high-fidelity qubit operations. At the same time, the switching between signal paths using transmon has been demonstrated~\cite{hoi2011demonstration}. This single-photon router shows high isolation efficiency ($\sim$ 92\%) with fast switching ($\sim$ few ns.), but with small bandwidth ($<$50~MHz), which makes it difficult to be widely used for multi-qubit scaling.
Even more  flexibility in EM-field steering is provided by two-input-two-output switches. The state-of-the-art implementation~\cite{pechal2016superconducting} gives high isolation and fast switching but it has a narrow bandwidth up to approximately $150~$MHz. That realization requires many on-chip components, including pairs of fixed-frequency hybrid beamsplitters, frequency-tunable resonators with chains of SQUIDs  and two DC-lines to control the switching state. As a result, this device has a large size of $5 \times 4$~mm, making  it difficult to use widely in on-chip multi-qubit processors. In addition, all of the above switches have fixed operating frequency.

\begin{figure*}
    \centering
    \includegraphics[width=1\linewidth]{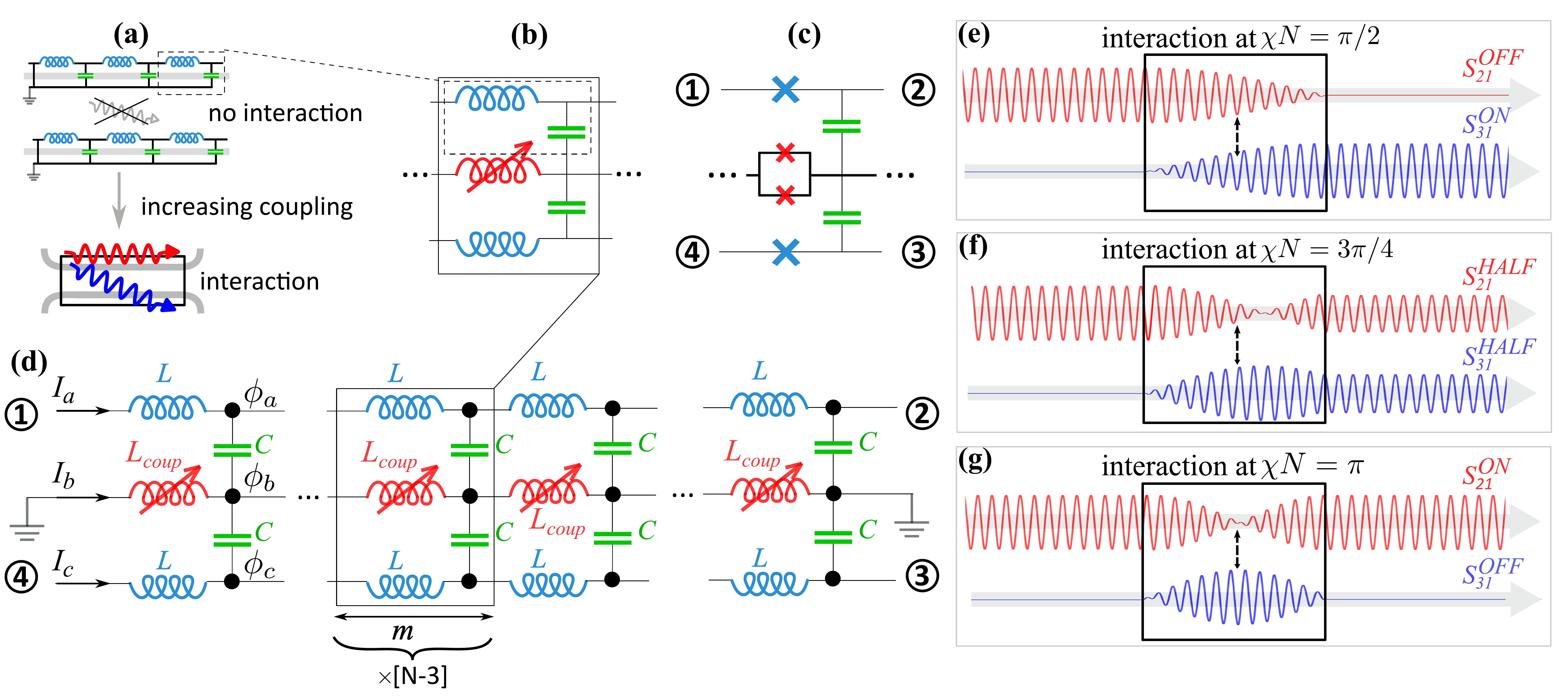}
    \caption{\textbf{Electric circuit and operating principle.} \textbf{(a)} A schematic of the four-port router with tunable coupling. \textbf{(b)} A lumped-element schematic realization of \textbf{(a)}. \textbf{(c)} A realization of the scheme \textbf{(b)} using compact components – Josephson junction inductors (blue), parallel-plate capacitors (green) and the tunable coupling as represented by SQUIDs (red). \textbf{(d)} The scheme used in the theoretical model. $\phi_a$, $\phi_b$, $\phi_c$ are potentials at the particular node, $I_a$, $I_b$, $I_c$ are current flows. \textbf{(e)-(g)} The cartoon of the three different regimes of switch states: $S_{21}^{OFF}$, $|S_{21}/S_{31}| = 1/2$, $S_{21}^{ON}$ based on the equation~\eqref{eq:IcoskappaN}. In the interaction box, the signal fully transmits to the other channel at different lengths due to a different coupling strength $\chi N$.}
    \label{idea}
\end{figure*}

Alternative approach to signal routing is based on two coupled transmission lines. Fixed coupling has been used for a directional coupler~\cite{pozar2009microwave} and microstrip-based networks~\cite{goldstein2022compact}. In microwave photonics, only tunable coupling between resonators, not transmission lines,  has been realized experimentally~\cite{wulschner2016tunable}, showing the narrow bandwidth ($\sim$~50~MHz) with low (1$<$GHz) frequency tunability. 
Although, in optical photonics, a switch based on tunable coupling between two suspended waveguides has been demonstrated~\cite{papon2019nanomechanical}. The coupling is controlled by changing the physical distance between the lines by external voltage and cannot be varied over a wide range. Also, this reason leads to a significant total length of the coupling section, which  reaches 20~times of the operating wavelength of $\lambda \sim 930$~nm. For microwave photonics with an operating frequency of 6~GHz on a silicon chip with $\varepsilon_{Si} \sim 12$, it implies structure length with the same  coupling strength, making direct realization of such an approach in microwave range impractical. 

%Another approach of a signal distribution is used by directional couplers, which  are based on two transmission line with fixed coupling between them. It has a wide frequency band, however it has a huge footprint and fixed coupling between transmission lines, that not allowed to use it as a switch. However, tunable, strong enough coupling strength between  transmission lines can merge broadband frequency range and switching performance. 
%The simple approach for routing devices could be a system, consisting of two transmission lines with tunable, strong enough coupling strength. In microwave photonics, there is also widely used directional couplers~\cite{pozar2009microwave} based on the same idea. However, it consist of waveguides on-chip, so their frequency properties and coupling strength cannot be tuned and determined by their physical size.
%In contrast to microwave optics, there is an experimental demonstration of this idea in photonic circuits. A switch is based on a tunable coupling between two suspended waveguides~\cite{papon2019nanomechanical}, which are controlled by changing physical distance between lines under external voltage. 

As one can see, there is a trade-off between sufficiently strong coupling, a reasonable structure size and coupling tunability. At the same time, in microwave photonics, there is no experimental realizations of a broadband on-chip two-input-two-output switch with frequency tunability suitable for signal routing. Moreover, we would like to make our device compact, i.e., smaller than the wavelength in coplanar waveguides at GHz frequencies ($\lambda_{cpw} = 10-40~$mm). We show that it is possible to solve all these problems at once by reducing the wavelength of the signal passing through the switch ($\lambda_{switch} = 0.1-0.4~$mm). This is achieved by choosing proper L and C elements forming the lumped-element transmission lines of the switch. 
%To achieve complete switching between transmission lines, the coupling strength between them must be strong enough. For the photonic circuits implementation, the total coupling length is distributed over distance, that is in 20~times larger than the operating wavelength $20\lambda$. For microwave photons with operating frequency of 6~GHz, it implies $20 \times 2$~cm = 40~cm structure length with the same mechanical coupling strength, making direct realization of such devices in microwave optics unrealistic.
%On the one hand, for optical photons it is hard to realize much stronger coupling -- it requires several nanometers of the gap between transmission lines. On the other hand,  for microwave photons, it is easy to realize strong enough coupling, such as for balanced hybrid beamsplitter~\cite{pozar2009microwave}, but without the possibility of tuning. This may be one possible reason why two-lines scheme has been realized in microwave optics only for fixed coupling strength as a beamsplitter~\cite{goldstein2022compact}. It shows a trade-off between sufficiently strong coupling strength (with reasonable structure size) and coupling tunability. At the same time, there is no experimental realization of tunable compact switch in microwave photonics.

The device has no moving parts, requires no pumping, and does not use standing waves, eliminating the size limitation of the structure. The device achieves high isolation more than 40~dB, for a 20~dB threshold the broadband range is several hundreds of MHz. The insertion loss is less than a few dB. The impedance change due to full-range tuning is within ~0.2~$\Omega$. The maximum operating power on the chip is -80~dBm,  corresponding to $2.5\times 10^6$ photons at 6~GHz per microsecond,  the highest maximum operating power to our knowledge for Josephson junction based devices. Also, the device demonstrates the  tunability over a frequency range of 4.8~--~7.3~GHz with above-mentioned  isolation quality.  The footprint of the device is $80\times420~\mu$m and it requires only a single DC coil for control by external flux bias. The device dissipates a negligible amount of heat during operation, and does not increase the temperature of the mixing chamber stage of a cryostat. The device is fabricated using UV lithography and requires only e-beam equipment for Josephson junction and parallel-plate capacitor fabrication~\cite{zotova2023compact}. This switch could provide flexible routing of electromagnetic fields, facilitate multiplexing and calibration, and help to scale up multi-qubit processors or miniaturize the current ones.  

\section{MODEL}

The switch consists of two transmission lines  with controllable coupling strength between them, as shown in Fig.~\hyperref[idea]{\ref*{idea}(a)}.  On the superconducting platform there are many ways to realize both the transmission lines and the tunable coupling between them by configurable inductance and capacitance, as shown in Fig.~\hyperref[idea]{\ref*{idea}(b)}. The inductance could be geometrical (a wire), kinetic (realized as thin films) or a Josephson inductance (represented by a Josephson junction). The capacitors could be either  interdigital (large) or  parallel-plate (compact). Tunable coupling could be realized by using variable capacitance~\cite{papon2019nanomechanical} or inductance,  represented by kinetic inductance or a chain of SQUIDs, which could be biased by the external magnetic flux or local flux bias. 

Next, we present a theoretical model describing the operation of the device. We denote the inductance of a single transmission line of each unit as $L$ and its capacitance as $C$. The coupling is represented by the mutual inductance of the lines, provided by the SQUID chain, $L_{coup}$ per unit. For the given operating angular frequency $\omega = 2 \pi f$, we find a relationship between the above parameters and the required total number of units $N$ to achieve a requested splitting ratio. The defined parameters and numbered ports are illustrated  in Fig.~\hyperref[idea]{\ref*{idea}(d)}. 

The system of wave equations describing the circuit in the continuous limit is 
\begin{equation}\label{eq:wave_equation}
\begin{cases} (1+\frac{L_{coup}}{L})\ddot{I}_a(n,t) + \frac{L_{coup}}{L}\ddot{I}_c(n,t) =   \frac{1}{CL}\frac{d^2 I_{a}(n,t)}{d n^2} \\ 
(1+\frac{L_{coup}}{L})\ddot{I}_c(n,t) + \frac{L_{coup}}{L}\ddot{I}_a(n,t) =  \frac{1}{CL}\frac{d^2 I_{c}(n,t)}{d n^2}. \end{cases}
\end{equation}Using Fourier transform of the current 
\begin{equation}
    I_{a,c} (k, \omega) = \int_{-\infty}^{+\infty} e^{j(kn - \omega t) }I_{a,c}(n, t) dndt
\end{equation} the wave equations~\eqref{eq:wave_equation} become: 
\begin{equation}\label{eq:after_Fourier}
\begin{cases} I_a(k, \omega)((1+\frac{L_{coup}}{L})\omega^2 - \frac{k^2}{LC}) + I_c(k, \omega)\frac{L_{coup}}{L}\omega^2= 0\\
I_a(k, \omega)(\frac{L_{coup}}{L})\omega^2 + I_c(k, \omega)((1+\frac{L_{coup}}{L})\omega^2 - \frac{k^2}{LC})= 0.
\end{cases}
\end{equation} 
This  equation~\eqref{eq:after_Fourier}  gives us two modes with the following dispersion relations for wave vectors 
\begin{align}\label{eq:disp_relat} \begin{split}
&k_+ = \pm \omega \sqrt{1 + 2\frac{L_{coup}}{L}}\sqrt{LC}\\
&k_- = \pm \omega \sqrt{LC}. \end{split}
\end{align}
When only the transmission line $a$ is excited at the input, the solution to the equation~\eqref{eq:wave_equation} is 
\begin{equation}\label{eq:currents} \begin{split}
\begin{pmatrix}
I_a \\
I_c 
\end{pmatrix}(n, t) \sim \begin{pmatrix}
e^{j(k_-n-\omega t)} + e^{j(k_+n-\omega t))} + c.c.\\
-e^{j(k_-n-\omega t)} + e^{j(k_+n-\omega t))} + c.c.
\end{pmatrix} &\sim \\
\sim \begin{pmatrix}
\text{cos}(Kn-\omega t)\text{cos}(\chi n)\\
-\text{sin}(Kn-\omega t)\text{sin}(\chi n) 
\end{pmatrix},\end{split}
\end{equation}
where  $K = 0.5(k_- + k_+)$ and $\chi = 0.5(k_- - k_+)$ show how the current oscillates: $K$ is the high-frequency carrier signal and $\chi$ is the low-frequency envelope. The phase factors of the currents of outputs 2 and 3 is equal to $cos(Kn-\omega t)$ and $-sin(Kn-\omega t)$, see eq.~\eqref{eq:currents}, so the phase difference between them is equal $-\pi/2$ and does not depend on the magnetic field.

For powers in output ports $S_{21}$ and $S_{31}$,  after averaging out over time  the high-frequency component $K$,  only the envelope $\chi$ for the oscillating current  is left 
\begin{align} \label{eq:IcoskappaN} \begin{split}
    &S_{12}\sim \langle I^2_a(N) \rangle \sim \text{cos}^2(\chi N),\\
    &S_{13}\sim \langle I^2_c(N) \rangle \sim \text{sin}^2(\chi N). \end{split}
\end{align}
The phase of the envelope signal determines how much of the signal passes to another channel. For the fully transmitted signal, see Fig.~\hyperref[idea]{\ref*{idea}(e)}, $I_a~=~0 \Rightarrow $  cos$(\chi N) = 0$ and $I_c = I_{max} \Rightarrow \text{sin}(\chi N) = \pm 1$, so the coupling phase
\begin{equation}\label{eq:kappaN}
    \chi N = \frac{1}{2}\left(\sqrt{1+2\frac{L_{coup}}{L}}-1\right)\sqrt{LC}\omega N = \frac{\pi}{2}.
\end{equation}
For other cases, such as equal splitting (Fig.~\hyperref[idea]{\ref*{idea}(f)}) $\chi N = 3\pi/4$, and return transmission to the first channel (Fig.~\hyperref[idea]{\ref*{idea}(g)})  $\chi N = \pi$.  Note, that only phase of the envelope signal at the output is important for the switching state, regardless of the number of oscillations passing through the system on the coupling length. Since the device is horizontally symmetric, and the coupling between the lines is described by the scalar parameter -- the inductance --  we expect the same behavior when we swap between inputs 1~and~4 and between outputs 2~and~3. If we apply signals to both inputs simultaneously, due to the linearity of the device, we expect the result to be equal to the sum of the corresponding complex amplitudes at the outputs, see eq.~\eqref{eq:currents}.

\begin{figure*}
\includegraphics[width=1\linewidth]{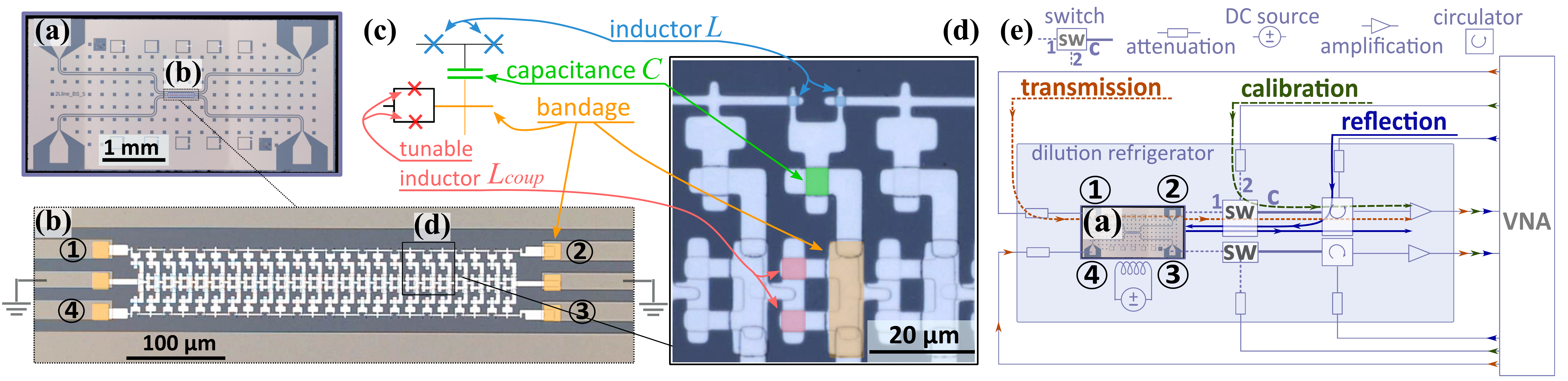}
\caption{\textbf{The chip and the measurement scheme.} \textbf{(a)} The micrograph of the chip, with enlarged parts \textbf{(b)} and \textbf{(d)} and the equivalent electric scheme \textbf{(c)}. The Josephson junction inductors (blue) and the capacitors (green) of each transmission line unit are shown.  The Josephson junction inductors, forming SQUIDs, are shown in red. Beige squares indicate bandages for robust galvanic contact. \textbf{(e)} The  measurement setup with the dilution refrigerator. Different colors of arrows indicate the path for the different measurements routes: transmission along the structure (dark orange short dashed line), reflection (dark blue solid line) and calibration channel (dark green long dashed line).  Filters, isolators and distribution of microwave components across refrigerator stages are omitted. The state of the commercial switch (~\textbf{\textquote{"sw"}}) is different for transmission-reflection~\textbf{\textquote{1}} and calibration measurement~\textbf{\textquote{2}} with common~\textbf{\textquote{c}} output. The setup is horizontally symmetrical along the chip.}
\label{chipmain}
\end{figure*}

The desired parameters  of the device, which fulfill the conditions~(\ref{eq:kappaN}) and impedance match $Z = \sqrt{L/C} = 50~\Omega$, are chosen taking into account the fabrication considerations, which we will discuss further. The choice of a particular inductor realization may require further refinement of the equations. In this paper, we present only a compact realization of the scheme, see Fig.~\hyperref[idea]{\ref*{idea}(c)}. To implement the inductors, we use Josephson junctions, which have the internal capacitance. The line inductance $L^*$ and the coupling inductance $L_{coup}^*$ including the effect of the Josephson junction capacitance are given by 
\begin{equation}\label{eq:JJcap}\begin{cases}
    &1/L^* = 1/L - \omega^2C_{JJ}\\
    &1/L_{coup}^* = 1/L_{coup} - \omega^2C_{SQUID}.\\ \end{cases}
\end{equation}

The inductance of each Josephson junction in the lines could be described by $L \approx \Phi_0/(2\pi I_c)$ and inductance of each SQUID as $L_{coup} \approx \Phi_0/\left(4\pi I_c \sqrt{\text{cos}(\pi \Phi/\Phi_0)}\right)$, where $\Phi_0$ is the magnetic flux quantum. These analytical expressions are valid as long as  the currents flowing through the junctions are small compared to  their critical values $I_c$. This leads to the limit in the working power, which we discuss in the next section. 

\section{Device realization}\label{MEASUREMENT SETUP AND DEVICE PARAMETERS}

In this section we present chip parameters and low-temperature measurement results. The chip and the equivalent electric scheme of the circuit are shown in Fig.~\hyperref[chipmain]{\ref*{chipmain}(a)-(d)}. Coplanar $50~\Omega$ waveguides for signal routing and flux traps for stabilizing magnetic field are formed on a silicon chip of 5~$\times~$2.5~mm. They are made of sputtered 50~nm niobium film followed by reactive ion etching in CF$_4$. Josephson junctions (blue and red) and capacitors~\cite{zotova2023compact} (green) consist of aluminum -- aluminum oxide -- aluminum trilayer in two separate lithography runs with thicknesses of 50~--~80~nm each. In order to provide a robust galvanic contact between several metallic layers, bandages (beidge) are used. They are made of 100~nm thick aluminum film with  heavy argon milling  prior to deposition to remove native oxidation layers of aluminium and niobium.  All aluminum structures are deposited using electron beam evaporation and oxidation in Plassys MEB550S3. Pattering is done by standard photolithography techniques and all aluminum structures are fabricated by lift-off. The area of the transmission line Josephson junctions is $1.8~\mu \text{m}^2$, the area of the capacitor is $22~\mu \text{m}^2$, the area of the SQUID is $26.4~\mu \text{m}^2$, with their junction areas of $12.6~\mu \text{m}^2$ each. The entire structure, consisting of $N=24$ units, has the size of $80~\times~$420~$\mu$m. To achieve a mirror-symmetric structure, we retain the total inductance of the Josephson junction of the edged unit in the transmission line but divide it into two Josephson junctions at the left and right edges. We realized this by changing the junction area; also, we omit the edged SQUID as it is shunted. We note that we also implemented the switches using large components (a wire inductor with  $S_L \sim 7600~\mu \text{m}^2$ and an interdigital capacitor with  $S_C \sim 34000~\mu \text{m}^2$). We find that for large components transmission is too inhomogeneous and has poor isolation. It can probably  be due to irregular ground potential over several millimeters,  crosstalk, or because the size of each unit is not too small in comparison to the operating wavelength. Due to the poor performance of a large-component device, we present only a compact realization and we emphasize the need for compact components, especially capacitors~\cite{zotova2023compact}.

\begin{figure*}[!]
    \includegraphics[width=1\linewidth]{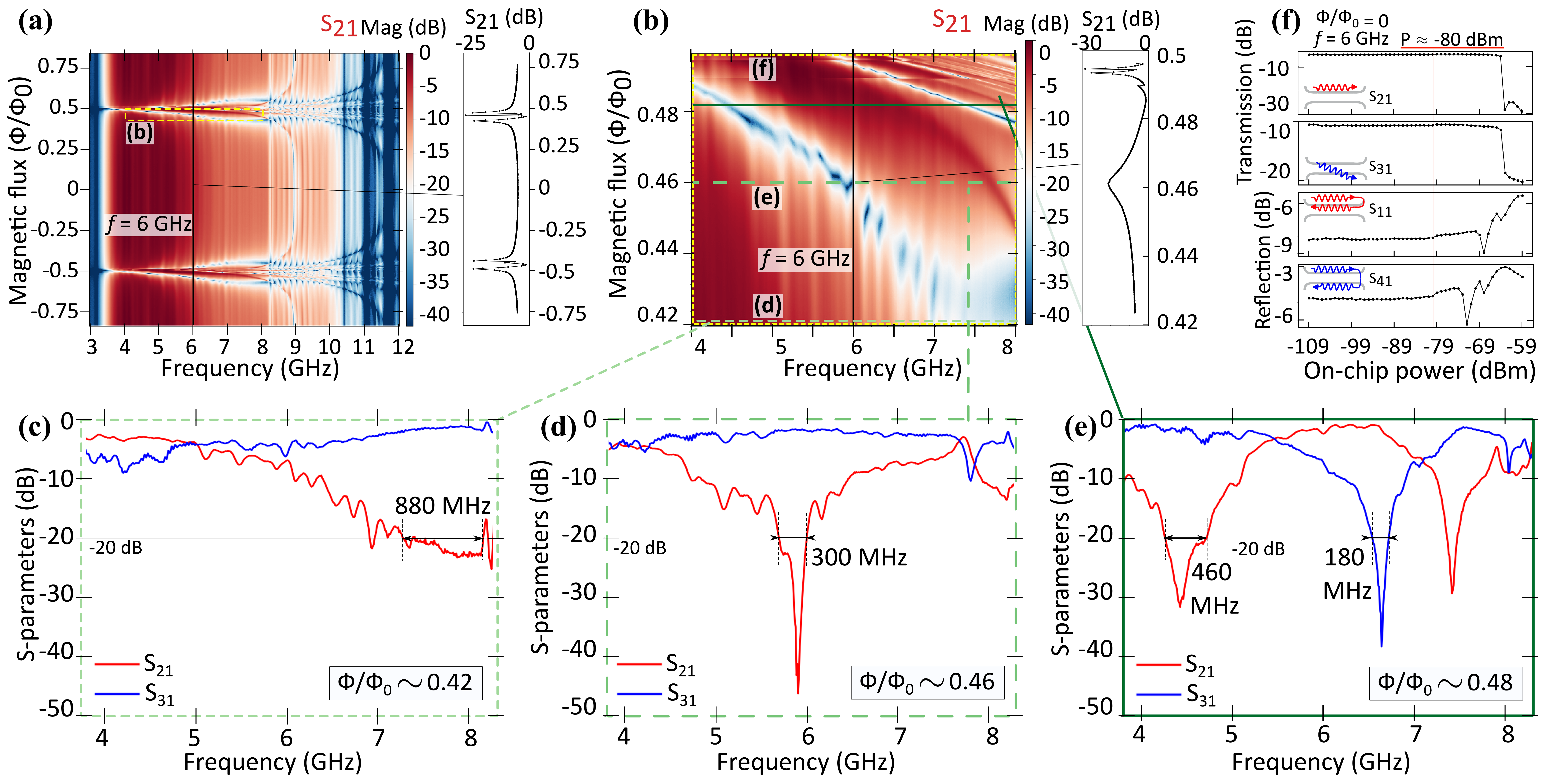}	
    \caption{\textbf{Selection of working  magnetic field and power.} \textbf{(a)} Magnitude of complex transmission $|S_{21}|$ of the device as a function of frequency and magnetic flux with a cross section at 6~GHz in the separate box (black), \textbf{(b)} enlarged area of a fragment of the periodic pattern near $\Phi/\Phi_0 = 0.5$ with cross sections at 6~GHz in the separate box (black) with cross sections \textbf{(c)}-\textbf{(e)}.  \textbf{(c)}-\textbf{(e)}: $|S_{21}|$ and $|S_{31}|$ as a function of frequency for different magnetic fields. The black horizontal line indicates the -20~dB isolation threshold. \textbf{(f)} Magnitude of raw complex transmission and reflection coefficients of ($|S_{21}|, |S_{31}|, |S_{11}|, |S_{41}|$) as a function of applied power at 6~GHz at $\Phi/\Phi_0 = 0$. The red vertical line is a threshold (-80~dBm on-chip) for the linear response.}
    \label{fluxpowercalib}
\end{figure*}

Using test Josephson junctions and bandages, placed on the same chip, we check their electric characteristics by measuring room-temperature resistance $R'_n$. Using the Ambegaokar-Baratoff relation $I_c = \Delta \pi/(2eR_n)$, where $R_n$ is a normal state resistance, $\Delta$ is the superconducting gap we estimate the inductance of each Josephson junction $L= \Phi_0/(2 \pi I_c)$. We note that $R_n < R'_n$, but for estimation we use $R_n = R'_n$ and get $L = 0.28~$nH/unit. 

For the resistance measurement, we consider the reference resistance of the bandage structure with its interface, $R \sim 42~\Omega$. For capacitors, we estimate the capacitance value as $C = S\times c$ of each unit as a product of the fabricated area $S$, obtained using an optical microscope and the capacitance per area value $c$, that we have obtained experimentally under similar conditions~\cite{zotova2023compact}. We find that the capacitance per each unit circuit  is $C \sim 300~$fF/unit. According to our estimates, the stray capacitance is  negligible compared to that value. Using parameters $L$, $C$ and the linear size of the unit $m = 34~\mu$m, we estimate the characteristic impedance $Z = \sqrt{L/C} \sim 30~\Omega$,  a constant part of the phase velocity $v_{ph}^{const} = m\omega/k_- = m/\sqrt{LC} = 3.4\times10^6~$m/s, and a variable part of it
\begin{equation}\label{eq:variable_vph}
    v_{ph}^{var} = m\omega/k_+ = m/ \left( \sqrt{1 + 2\frac{L_{coup}}{L}}\sqrt{LC}\right),
\end{equation}
which we numerically estimate in the end of the measurement section after fitting the measurement data. The slow phase velocity is the key feature of this transmission line, since it limits the spatial size of the device. In the case of using conventional transmission line, the device will be large as the phase velocity faster, that is more than 5~mm long.

\section{Measurement results}
The chip is mounted at the base stage of a dilution refrigerator at temperature of 12~mK. The signal from a vector network analyzer (VNA) is attenuated and filtered before entering the chip and amplified at the output. The raw magnitude of the complex transmission coefficients $|S_{21}|$ and $|S_{31}|$ parameters are obtained under continuous wave excitation, generated by the VNA, and then calibrated by the separate path in the cable setup, see the measurement scheme in Fig.~\hyperref[chipmain]{\ref*{chipmain}(e)}. The measurement setup inside the dilution refrigerator consists of two nominally identical sets of an input line for the transmission measurement (dark orange short dashed line), an  input line and a circulator for reflection measurement (dark blue solid line) and a line for calibration (dark green long dashed line). These three paths are connected by a commercial switch, and one can select transmission, reflection or calibration using the same source of microwave signals. Calibration takes into account the path from the commercial switch to the output, therefore, cables, connectors of the sample holder, bonds between the chip and a commercial switch affect the calibrated results by about a few~dB. We present the results in a range of 4~--~8~GHz limited by our circulator bandwidth. When the fourth port is not used, it is terminated with a 50~$\Omega$ load.

\begin{figure*}
    \centering
    \includegraphics[width=1\linewidth]{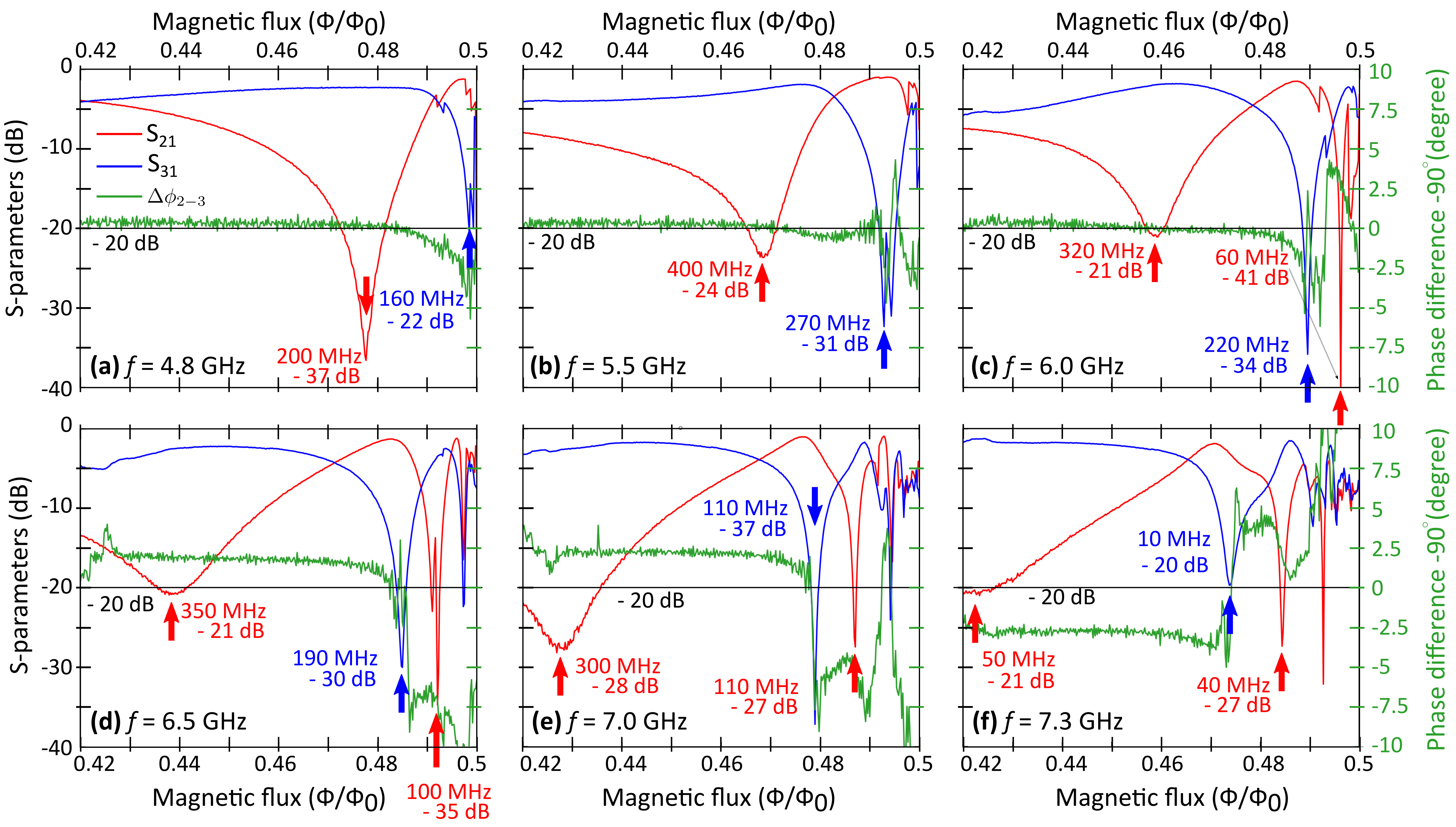}
    \caption{\textbf{Switching performance for different frequencies.} Calibrated magnitude of complex transmissions $|S_{21}|$ (red) and $|S_{31}|$ (blue) for each frequency with varying magnetic flux, generated by  the global coil.  The black horizontal line shows a -20~dB level, the red and the blue arrows indicate possible working points of switching to the requested state. The bandwidth and isolation are also shown in these points. The green line shows the phase difference between output ports 2 and 3.}
\label{xplot}
\end{figure*}

First, we obtain the flux-dependent response of the S-parameters (Fig.~\hyperref[fluxpowercalib]{\ref*{fluxpowercalib}(a),(b)} for $|S_{21}|$). We measure $|S_{21}|$ and $|S_{31}|$ as a function of frequency and external magnetic flux applied by the coil.  We observe a periodic pattern, which is expected from the periodic behavior of the Josephson energy and inductance of SQUIDs on magnetic flux.  One period of the pattern of $|S_{21}|$ is shown in Fig.~\hyperref[fluxpowercalib]{\ref*{fluxpowercalib}(a)} with a vertical slice of the data at 6~GHz in black. The transmission changes a little from $\Phi/\Phi_0 = 0$ to almost $\Phi/\Phi_0 = 0.4$, but changes radically  when approaching half-flux quantum. This enlarged part of the $|S_{21}|$ spectrum for $\Phi/\Phi_0 = 0.42 - 0.50$ is shown in Fig.~\hyperref[fluxpowercalib]{\ref*{fluxpowercalib}(b)} with the section at 6~GHz in black. The resonance peaks shift towards lower frequencies as the inductance of the coupling SQUIDs increases with increasing magnetic flux, as shown in Fig.~\hyperref[fluxpowercalib]{\ref*{fluxpowercalib}(b)-(e)}.  The area near $\Phi/\Phi_0 \sim 0.5$ has the sharpest transmission changes, where the inductance of each SQUID undergoes the largest changes and the maximum value is limited only by the asymmetry of the Josephson contacts. Fig.~\hyperref[fluxpowercalib]{\ref*{fluxpowercalib}(c)-(e)} shows calibrated $|S_{21}|$ (red) and $|S_{31}|$ (blue) transmissions for different magnetic fluxes $\Phi/\Phi_0 = 0.42$ (Fig.~\hyperref[fluxpowercalib]{\ref*{fluxpowercalib}(c)}, short green dashed line), $\Phi/\Phi_0 = 0.46$ (Fig.~\hyperref[fluxpowercalib]{\ref*{fluxpowercalib}(d)}, long green dashed line) and $\Phi/\Phi_0 = 0.49$ (Fig.~\hyperref[fluxpowercalib]{\ref*{fluxpowercalib}(e)}, solid green line). The data show a broadband transmission of 0.2 -- 1~GHz for the operating regime with at least 20~dB of isolation, exceeding 40~dB for some frequencies. Transmission is slightly below the unit level, likely due to impedance mismatch and imperfect calibration procedure.

We also estimate the linear power operating range of our device, see Fig.~\hyperref[fluxpowercalib]{\ref*{fluxpowercalib}(f)}. At $\Phi/\Phi_0 = 0$ and  $f = 6~$GHz we plot the longitudinal $|S_{21}|$ and transverse $|S_{31}|$ transmissions and reflections  $|S_{11}|, |S_{41}|$ as a function of the on-chip power, which is estimated using calibration protocol, mentioned above.   We find that the device exhibits linear behaviour up to approximately $-80$~dBm, which is indicated by the red vertical line. This power is equivalent to 10~pW,  $2.5\times 10^{6}$ photons at 6~GHz per microsecond, which is sufficient not only for qubit and single-photon experiments, but also for high-power purposes, such as pumping and driving qubit systems. The sharp drop in the transmission $|S_{21}|, |S_{31}|$, around $-65$~dBm of applied power, corresponds to exceeding the critical current through  Josephson junctions in the transmission line, where the analytical expression for the inductance of the Josephson junction is not valid and active losses occur. In general, the upper limit of the operating power can be increased if the number of $N$ units is increased with successive increases in the critical current of each Josephson junction, keeping the total inductance of the SQUID chain the same. We assume that for other magnetic fields the device will still be in a linear regime below the threshold power, since the device linearity is determined by the linearity of the transmission line Josephson junctions. This is due to the fact that their critical current is 10 times smaller than that of SQUIDs.

Next, we discuss the switching quality. We plot $|S_{21}|$ and $|S_{31}|$ as a function of external normalized magnetic flux  at different operating frequencies in a working range of 4.8~GHz – 7.3~GHz -- 4.8, 5.5, 6.0, 6.5, 7.0, 7.3~GHz as shown in Fig.~\hyperref[xplot]{\ref*{xplot}}. These plots are vertical cut-plots of Fig.~\hyperref[fluxpowercalib]{\ref*{fluxpowercalib}(b)} for calibrated $S_{21}$ and $S_{31}$. As one can see, the curves of each panel of Fig.~\hyperref[xplot]{\ref*{xplot}} continuously change from one panel to the next as the frequency changes, giving us an understanding of what the S-parameters are like at other frequencies within the operating range. The isolation for $|S_{31}|$ and $|S_{21}|$ for each frequency  exceeds 20~dB, which is indicated by the black horizontal line in each subplot in Fig.~\hyperref[xplot]{\ref*{xplot}}. The possible operating points of the device are indicated by red and blue arrows, accordingly. At lower magnetic flux, the transmission minimum is shallower compared to stronger fields, but smoother and less sensitive to changes of magnetic field, such as noise caused by flux jumps or instability of a current source. Depending on the requirements – higher isolation or insensitivity to magnetic noise – different magnetic flux points should be chosen. Also, at $\Phi/\Phi_0 = 0.49 - 0.50$  at high frequencies we observe jumps in transmissions, that may be caused by inhomogeneities of critical currents in SQUIDs, forming coupling between transmission lines. Also, the data demonstrates  the device can operate as a beamsplitter with splitting of equal or an arbitrary ratio and low $<$~1 dB loss, although the bandwidth is narrow. As for the measured phase difference between the outputs, we cannot precisely measure the absolute value of the phase difference, but we have observed a constant value as a function of the magnetic flux, see Fig.~\hyperref[xplot]{\ref*{xplot}}~(green). The value is different for different frequencies, so the data are shifted to the theoretical reference of -90$\degree$  for clarity. The phase data shows a negligible spread of less than 10$\degree$ in the working range of the magnetic flux.

\begin{figure}
    \centering
    \includegraphics[width=1\linewidth]{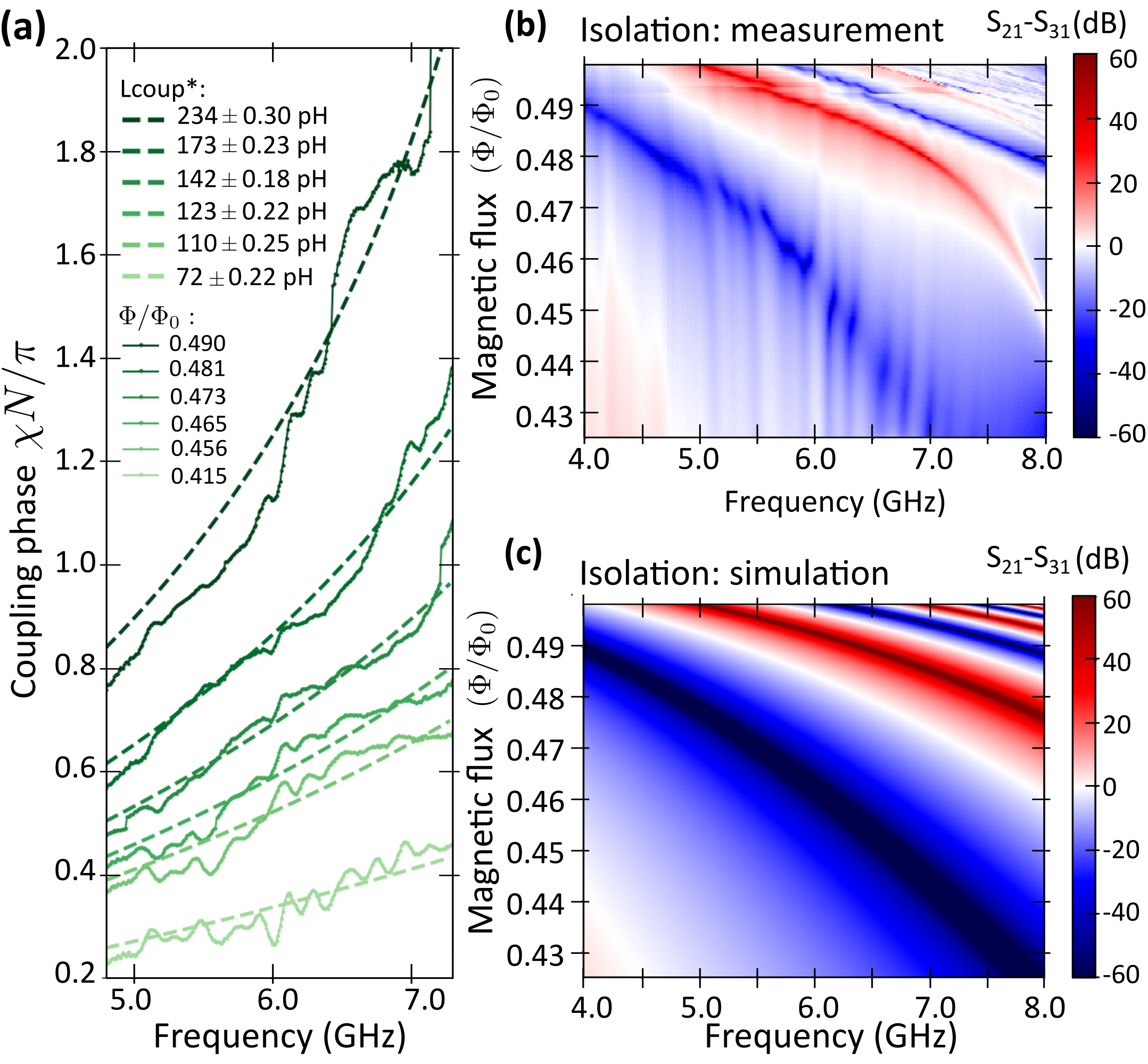}
    \caption{\textbf{Fitting of coupling phase and numerical calculations of the model.}  \textbf{(a)} The coupling phase between two transmission lines as a function of frequency for different magnetic fluxes. Dotted lines: data extracted using the formula~\eqref{eq:IcoskappaN}. Dashed lines: fitted using the formula~\eqref{eq:kappaN} and~\eqref{eq:JJcap}. \textbf{(b)}: isolation ratio as a function of frequency for different magnetic fluxes of the calibrated S-parameters, calculated as  $S_{21}-S_{31}$. \textbf{(c)}: Numerical calculation of the isolation ratio of the S-parameters (formulae~\eqref{eq:IcoskappaN} and ~\eqref{eq:kappaN}) as a function of frequency for different magnetic fluxes.}
\label{kappaN}
\end{figure}

Below we discuss the  fitting model for S-parameters and investigates the coupling phase between two transmission lines as a function of different magnetic fields.  Using the equation~\eqref{eq:IcoskappaN} we extract $\chi$  from the experimental data as a function of frequency for different magnetic fluxes, indicated by different green color intensity and plot the coupling phase $\chi N/\pi$ in Fig.~\hyperref[kappaN]{\ref*{kappaN}(a)}, dots. Here, we select the same frequency range as for the  switching quality demonstration -- 4.8~GHz -- 7.3~GHz. Next, we fit the data  using the equations~\eqref{eq:kappaN} and~\eqref{eq:JJcap} and extract the effective inductance $L_{coup}^*$ of each unit, see Fig.~\hyperref[kappaN]{\ref*{kappaN}(a)}, dashed lines. Using the extracted values of $L_{coup}^*$, we calculate the coupling phase of all other current values and use it for further numerical calculations. We plot the isolation ratio of the measured S-parameters $S_{21}-S_{31}$ as a function of frequency and applied external magnetic flux, see Fig.~\hyperref[kappaN]{\ref*{kappaN}(b)}.  The isolation ratio of the S-parameters is then calculated using the equations~\eqref{eq:IcoskappaN}, \eqref{eq:kappaN} and \eqref{eq:JJcap} and the fitting data from Fig.~\hyperref[kappaN]{\ref*{kappaN}(a)} as the same function of frequency and magnetic flux, see Fig.~\hyperref[kappaN]{\ref*{kappaN}(c)}.  We see, that the model describes the qualitative S-parameters well, with a small inhomogeneous discrepancy, which could be explained by the non-ideal identity of the coupling SQUIDs. The difference between the absolute numerical value of the calculated and the measured isolation ratio could be explained by non-zero parasitic reflections,  possible dielectric losses and RF crosstalk between the channels.  In general, the plots Fig.~\hyperref[kappaN]{\ref*{kappaN}(b),(c)} show that each frequency in the working range of 4.8~GHz -- 7.3~GHz has the maximum and the minimum value of the isolation ratio, corresponding to the open and closed states for the each output of the switch.
In addition, increasing the magnetic field, and hence the coupling inductance, results in decrease in the variable component of the phase velocity [Eq.~\eqref{eq:variable_vph}] $v_{ph}^{var}$ from $2.75\times 10^6$ m/s to $2.07\times 10^6$ m/s, and the characteristic impedance $Z$ by 0.5\%  according to $Z = 0.5(\sqrt{1 + 2L_{coup}/L} + 1)\sqrt{L/C}$.

Finally, we discuss the characteristic time of switching between the channels and changing operating frequencies. The presented device is controlled by the external magnetic field, generated by a coil with a large coil inductance. In this case the switching time is limited by the characteristic rise time of the coil. 
To make the switching time shorter, high-frequency current bias lines should be implemented on-chip similar to the ones used to control energies of superconducting qubit with SQUID loops. The pulses in the matched broad-band line are limited by  tens of picoseconds. In practice, the switching time in this case  is limited by the speed of an arbitrary waveform generator. As a matter of fact, devices based on a SQUID chain can change their characteristics in  nanoseconds~\cite{sandberg2008tuning}. The minimal possible switching time is limited by the transient processes in each SQUID in response to the magnetic field, which we estimate as reversed Josephson junction plasma frequency $\tau_{JJ} = 1/\omega_{p} = \sqrt{L^*_{coup}\times C_{SQUIDs}} \sim 12$~ps. The switching time also depends on a propagation time along the coupling length: $\tau_{1-2} = mN/v_{ph}^{const} \approx 124~ps$ or $\tau_{1-3} = mN/v_{ph}^{var} \approx 101-76~ps$ depending on the magnetic field, where the phase velocity is determined by the formula~\eqref{eq:variable_vph}. The total exact switching time $\tau = \tau_{JJ} + \tau_{1-2}$ (or $\tau = \tau_{JJ} + \tau_{1-3}$) depends on the magnetic field, and for simplicity, we can determine the switching time by the longest delay as $\tau  = 136~ps$.

\section{Conclusion}
We describe the design, realization, and characterization of the switch, based on two transmission lines, coupled by a tunable inductor, implemented by a chain of SQUIDs. We experimentally demonstrate  that the operating frequency is tunable over a range  4.8~GHz~--~7.3~GHz with a bandwidth of a few hundreds of MHz. For a  given bandwidth, the isolation exceeds 20~dB and reaches  40~dB. The working power ranges up to ~$-80$~dBm on-chip, corresponding to $2.5\times 10^6$ photons per microsecond (10~pW) at 6~GHz frequency. The device size is $80 \times420~\mu$m and it is controlled by external flux bias with local DC-line control capability. The impedance change due to tuning is within 0.2~$\Omega$. The fabrication of the investigated device requires only well established fabrication techniques, such as  photolithography and e-beam deposition and oxidation. The switching time is limited by the external magnetic control, such as speed of the current change in a coil. Operating at the base temperature of a dilution refrigerator, the switch does not heat the circuit.

To improve the performance of the device, e.g. as to increase the working frequency range, and to avoid approaching to $\Phi_0/2$ due to flux jumps, the reasonable strategy is to increase the number of units $N$, while keeping other parameters the same.  Increasing $L_{coup}$ itself by reducing the critical current $I_c$ of the Josephson junctions forming the SQUIDs also leads to the same results, but at the same time, it causes a decrease in the working power limit.  Also, future work could aim at fast control using local bias, as this could  reduce the switching time. The bias lines could be implemented by using the air-bridge technology~\cite{chen2014fabrication}. In this case it will be possible to operate with multiple devices on the same chip, opening a door to many microwave multiplexing applications and configurable networks schemes. The device can be used to control and read out superconducting quantum systems with fewer number of cables. In addition, there is a possibility of operating the device in the quantum regime, by using a single-photon source~\cite{zhou2020tunable}. The device show a possibility to be operated as a mirror (or a beamsplitter) with tunable splitting ratio, and the future research can be dedicated to the investigation the efficiency of the splitting and applications in linear optic schemes.

\section{Acknowledgements} 
J.Z. is grateful to Hiroto Mukai, Ilya Besedin and Gleb Fedorov for stimulating discussions. This work was funded by RIKEN IPA Program,  ImPACT Program of Council for Science, Technology and Innovation (Cabinet Office, Government of Japan),  by CREST, JST  (Grant No. JPMJCR1676) and by the New Energy and Industrial Technology Development Organization (NEDO), JPNP16007. J.Z. is grateful to the support of the Russian Science Foundation (grant \#21-72-30026 https://rscf.ru/en/project/21-72-30026, data analysis). A. S. is grateful to the support of the Russian Science Foundation (grant \#21-72-10117, theory).

%\newpage
%\clearpage
\bibliography{ref}
\bibliographystyle{unsrt}
\end{document}